\documentclass[aps,prl,twocolumn,showpacs,groupedaddress]{revtex4}
\usepackage{amsmath,amsfonts,bm}
\usepackage{graphicx,epsfig}
\usepackage{color}

\usepackage{epstopdf}

\begin{document}

\title{Arrested coalescence of viscoelastic droplets: Triplet shape and restructuring}

\author{Prerna Dahiya,$^{1}$ Andrew DeBenedictis,$^{2}$ Timothy J. Atherton,$^{2}$
Marco Caggioni,$^{3}$ Stuart W. Prescott,$^{1}$ Richard W. Hartel,$^{4}$  and Patrick T. Spicer,$^{1}$}

\affiliation{$^{1}$Complex Fluids Group, School Chem. Eng., UNSW Australia, Australia,}
\affiliation{$^{2}$Soft Matter Theory, Physics \& Astronomy Depart., Tufts University, USA,}
\affiliation{$^{3}$Complex Fluid Microstructures, Procter \& Gamble Co., USA,}
\affiliation{$^{4}$Food Engineering Depart., University of Wisconsin-Madison, USA}

\begin{abstract}
The stability of shapes formed by three viscoelastic droplets during their 
arrested coalescence has been investigated using micromanipulation experiments.
Addition of a third droplet to arrested droplet doublets is shown to be controlled by 
the balance between interfacial pressures driving coalescence and internal elasticity
that resists total consolidation. The free fluid available within the droplets controls the
transmission of stress during droplet combination and allows connections to occur via
formation of a neck between the droplets. The anisotropy of three-droplet systems adds complexity
to the symmetric case of two-droplet aggregates because of the multiplicity
of orientations possible for the third droplet. When elasticity dominates, the initial
orientation of the third droplet is preserved in the triplet's final shape. When elasticity is dominated
by the interfacial driving force, the final shape can deviate strongly from the initial positioning of 
droplets. Movement of the third droplet to a more compact packing occurs, driven by liquid meniscus
expansion that minimizes the surface energy of the triplet.  
A range of compositions and orientations are examined and the resulting domains of 
restructuring and stability are mapped based on the final triplet structure.  A geometric and a physical model are used to explain the mechanism
driving meniscus-induced restructuring and are related to the impact of these phenomena
on multiple droplet emulsions.
 
\end{abstract}


\maketitle

\section{Introduction}
Emulsions are mixtures of two immiscible liquid phases, where one phase exists as 
dispersed droplets in the second, continuous, phase as long as the system is stabilized against coalescence: the 
recombination of the dispersed phase into a single liquid region.  Coalescence can be prevented 
by a number of mechanisms, but one common to many complex fluids like foods is the 
arrest resulting from colloids that provide an offsetting rheological resistance.\cite{Boode:1993vp,Dickinson:2012hka,Vanapalli:2001tz} 
Arrest occurs once coalescence is initiated but can not complete because of a 
halt in the ability to reduce droplet surface area \cite{Stratford:2005wb, Studart:2009to,Pawar:2011tq} 
or increase deformation\cite{Pawar:2012wj,Dahiya:2016hz} to minimize system energy. 
While both types of arrest play a role in practical microstructures, the surface, or 
Pickering, type can have significant limits on the number of connections formed 
between droplets\cite{Wu:2015ie} because the end result is essentially a solid interfacial shell. 
Droplets with an internal viscoelastic resistance are more flexible in their ability to form 
multiple connections between droplets, even after an initial arrest event. 

 \begin{figure}
  \centering
  \includegraphics[scale=0.35]{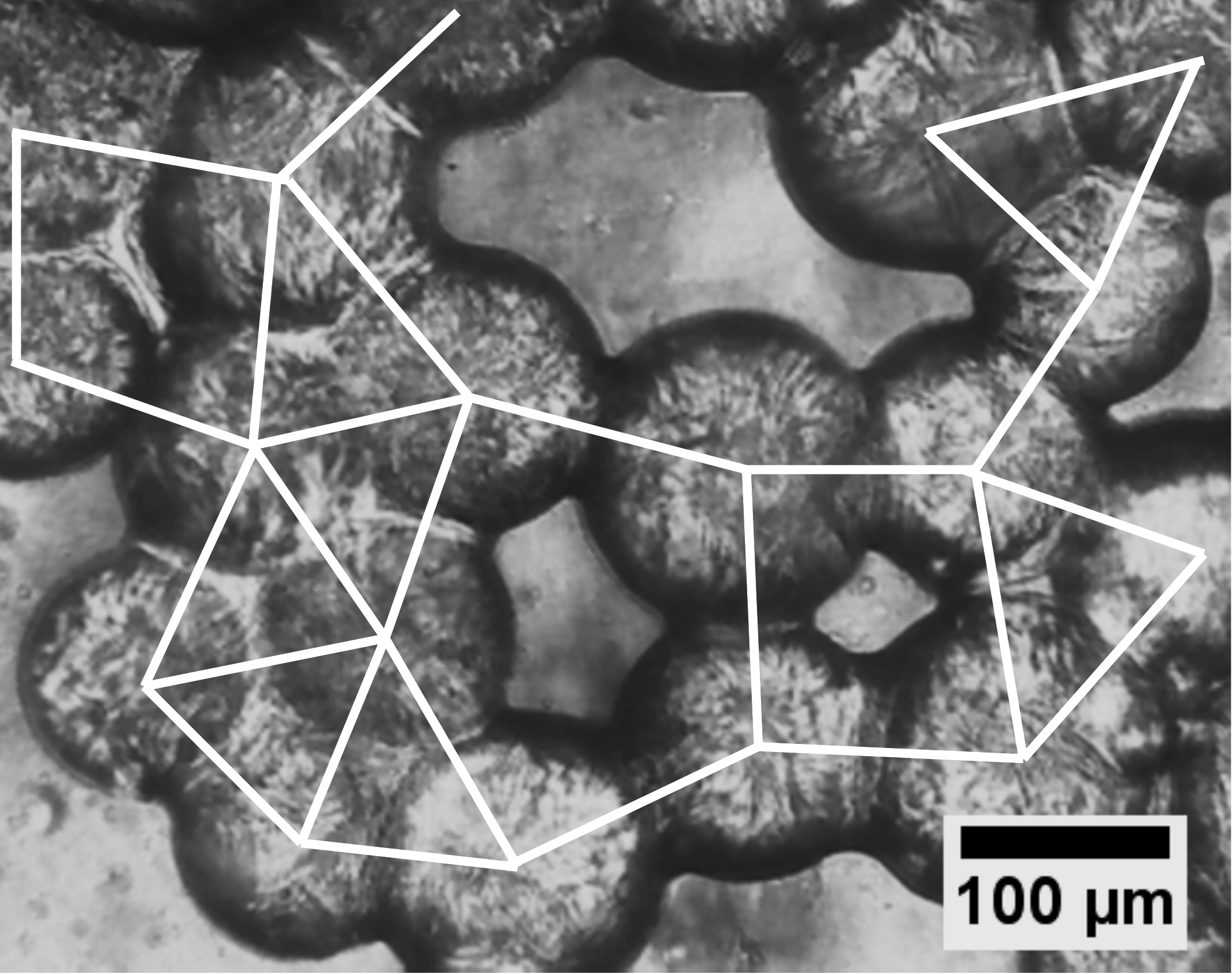}
  \caption{An emulsion exhibiting arrested coalescence of a number of droplets, 
  with the connections mapped by white lines. 
  A wide range of angles between 
  droplets is visible. Some droplets are quite closely packed 
  while others have a surprising amount of space between them.}
  \label{fig:multiangles}
\end{figure} 

Previous work showed that droplets containing an elastic network of crystals can initiate but not 
finish coalescence, creating a range of anisotropic intermediate droplet shapes,\cite{Boode:1993vp,Fredrick:2010tu} 
but only recently has a physical model of the process for uniform\cite{Pawar:2012wj} and non-uniform\cite{Dahiya:2016hz} 
droplet pairs been developed.  Essentially the interfacial pressure driving two 
droplets to minimize their area must be offset by the internal elasticity to stably arrest 
coalescence at an intermediate state: a liquid doublet rather than a sphere.\cite{Pawar:2012wj}   
In such cases the exact doublet shape represents one of a number of intermediate 
points between the beginning state  of two discrete
droplets and a completely coalesced single sphere and is determined by
the level of solids in the droplet and the point at which coalescence is arrested.\cite{Pawar:2012wj}   
We extend that work here by examining the dynamic behavior of an aggregate of three arrested droplets, a triplet,
 as multiple connections are the basis for many fluid microstructures formed during the 
 manufacture of food, coatings, and material templates like bijels.\cite{Mohraz:2016hd} 
 Unlike aggregated solid particles, however, arrested droplets are permeated by a liquid 
 oil phase that can adhere easily to other surfaces and can form a neck that bridges the two droplets.  
 It is this liquid bridge that determines connectivity 
 between droplets and, by transmitting interfacial pressures that balance the internal elasticity of the droplets, 
 the degree of deformation exhibited. When two droplets connect to form a doublet, the interfacial driving force for 
 coalescence acts along the center of each droplet and produces a rotationally symmetric shape. 
 When an additional droplet is added, however, that symmetry is broken and more complex behavior is expected.

These experiments report the dynamic behavior of three viscoelastic droplets arrested 
at an intermediate stage of coalescence by internal elasticity. 
Micromanipulation and microscopic time-lapse imaging techniques are used to study 
changes in droplet deformation and order to  investigate the path taken to a final 
droplet aggregate structure. We observe a significant degree of droplet restructuring 
from loose to dense packings as long as the droplet solids 
content and elasticity are low. We develop a physical model of the restructuring, 
resulting in a simple geometric criterion predicting the extent of densification as a function of system variables.

Our ultimate goal is to build on previous work on local deformations of structured emulsion 
droplets\cite{Pawar:2012wj,Dahiya:2016hz} to realistically simulate the formation of full-scale 
emulsion networks\cite{ThivilliersArvis:2010vi,thiel2016coalescence} and more complex coalescence 
phenomena.\cite{thiel2016coalescence} We increase the complexity of 
scrutiny here by studying triplets of structured droplets as these break the symmetry of the 
doublets previously examined, exhibiting new mechanisms of shape-change and construction.  
An understanding of the packing and connectivity within such structures is critical to predicting 
mechanical properties and dynamic performance of materials as diverse as foods, cosmetics, 
and 3D printed products.\cite{Thivilliers:2006tg,Thivilliers:2008vk,ThivilliersArvis:2010vi} This work 
also demonstrates a new aspect of responsiveness and shape-change: restructuring of an 
underlying elastic framework using the strong driving force of a liquid 
interface.\cite{Leong:2007uf,Py:2007ts,2010RomanBico,Caggioni:2014kw,Caggioni:2015kh,Prileszky:2016gc} 
We show that numerous complex shapes can be self-assembled even from simple droplet building blocks
larger than the thermal limit.

 \begin{figure*}
  \centering
  \includegraphics[scale=0.9]{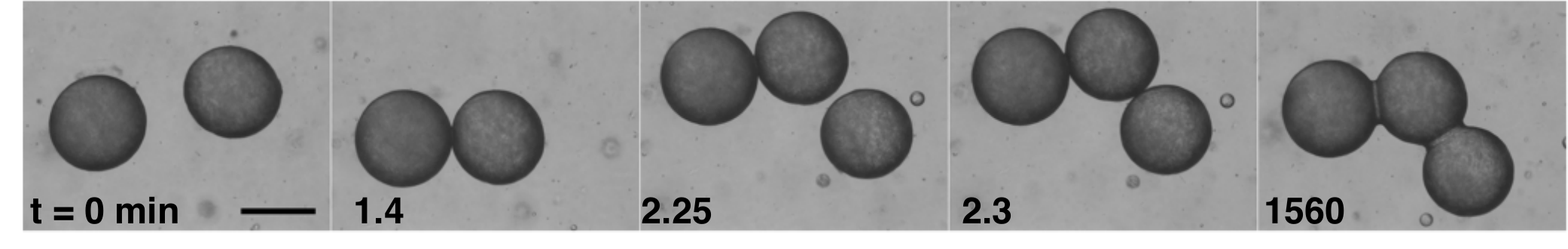}
  \caption{Successive images of arrested coalescence of droplets at 40\% solids level, showing 
  changes in area and compactness of the shape. Time of each image 
  acquisition is given in minutes. Scale bar is $200 \; \mu m$.}
  \label{fig:40restr}
\end{figure*}

 \begin{figure*}
  \centering
  \includegraphics[scale=0.9]{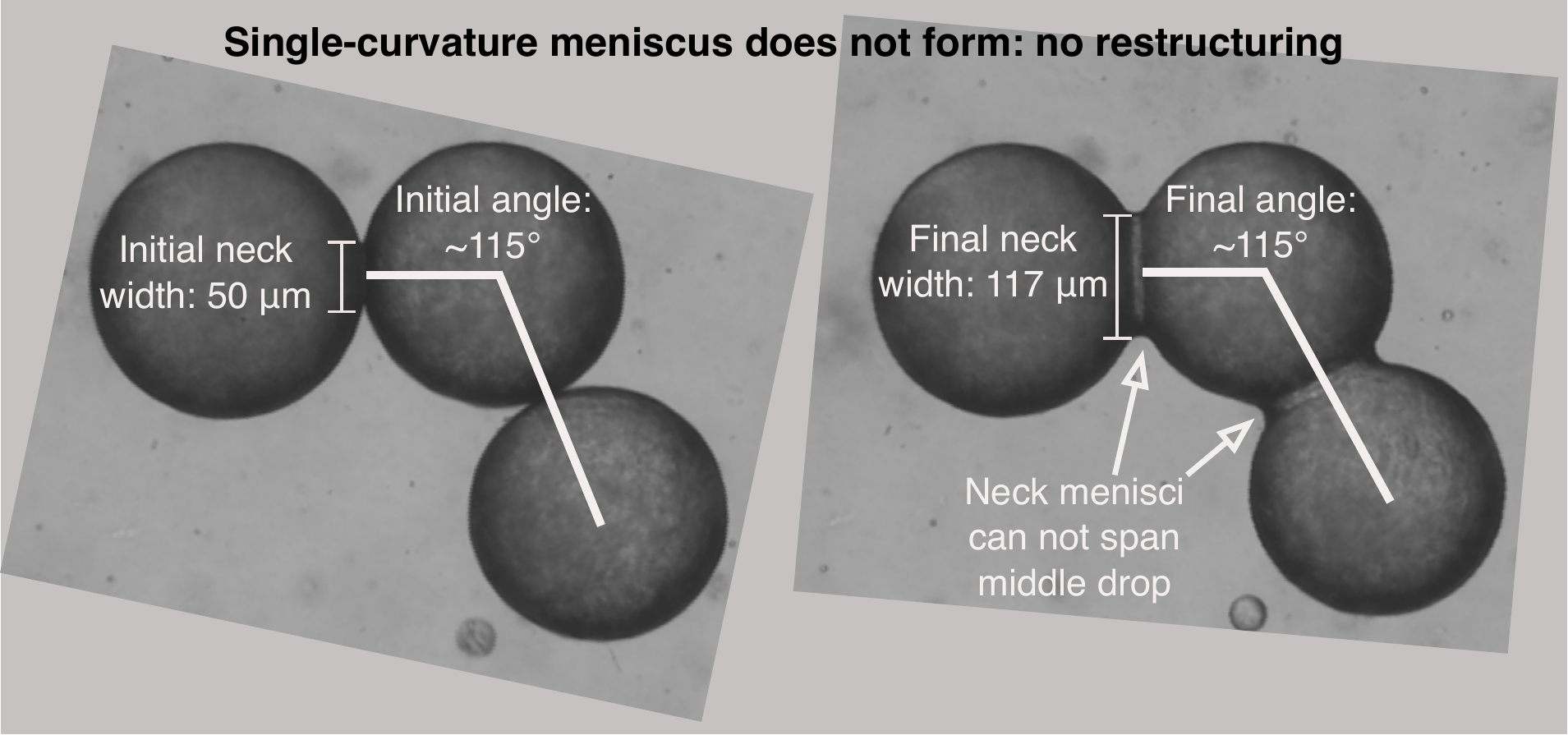}
  \caption{Close-up study of the changes occurring during triplet formation in 
  Figure \ref{fig:40restr}, showing neck width growth and two separate meniscus regions.}
  \label{fig:norestrschem}
\end{figure*}

 \begin{figure*}
  \centering
  \includegraphics[scale=0.9]{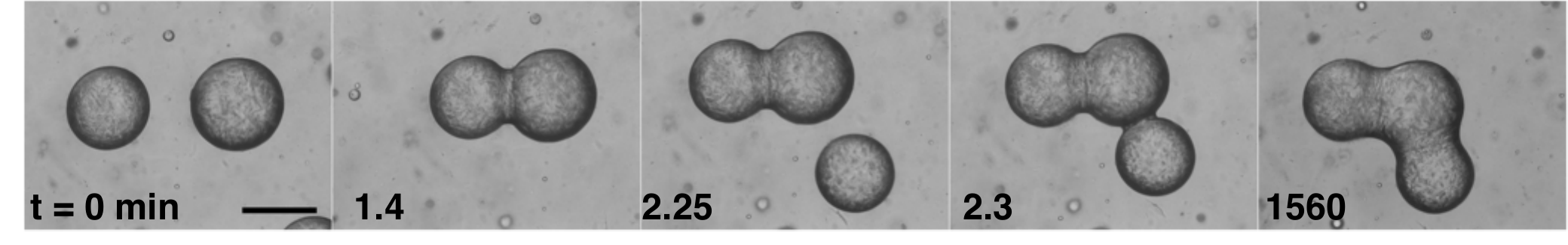}
  \caption{Arrested coalescence of droplets at 30\% solids level, where changes in 
  the meniscus between droplets can occur but are not strong enough to significantly restructure the triplet.
  Time in minutes, scale bar $200 \; \mu m$.}
  \label{fig:30restr}
\end{figure*}

 \begin{figure*}
  \centering
  \includegraphics[scale=0.9]{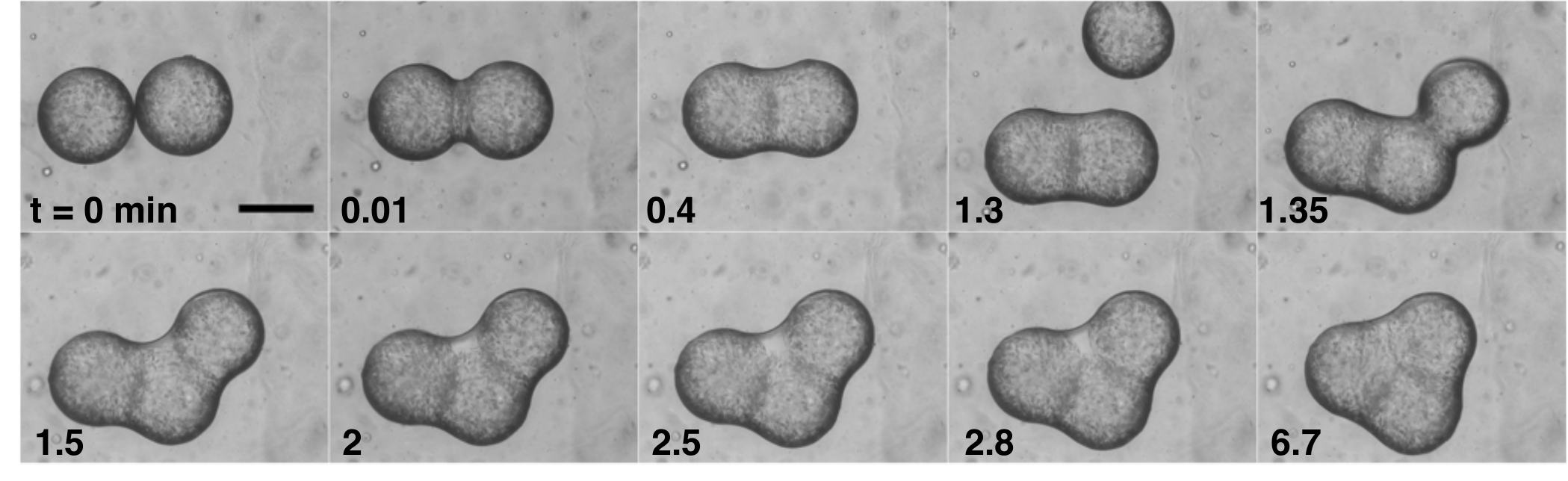}
  \caption{Images of restructuring of droplets (25\% solids) as a result of meniscus 
dynamics, the driving force to make the droplet relocate to a denser packing state. 
Time in minutes, scale bar $200 \; \mu m$.}
  \label{fig:25restr}
\end{figure*}

 \begin{figure*}
  \centering
  \includegraphics[scale=0.9]{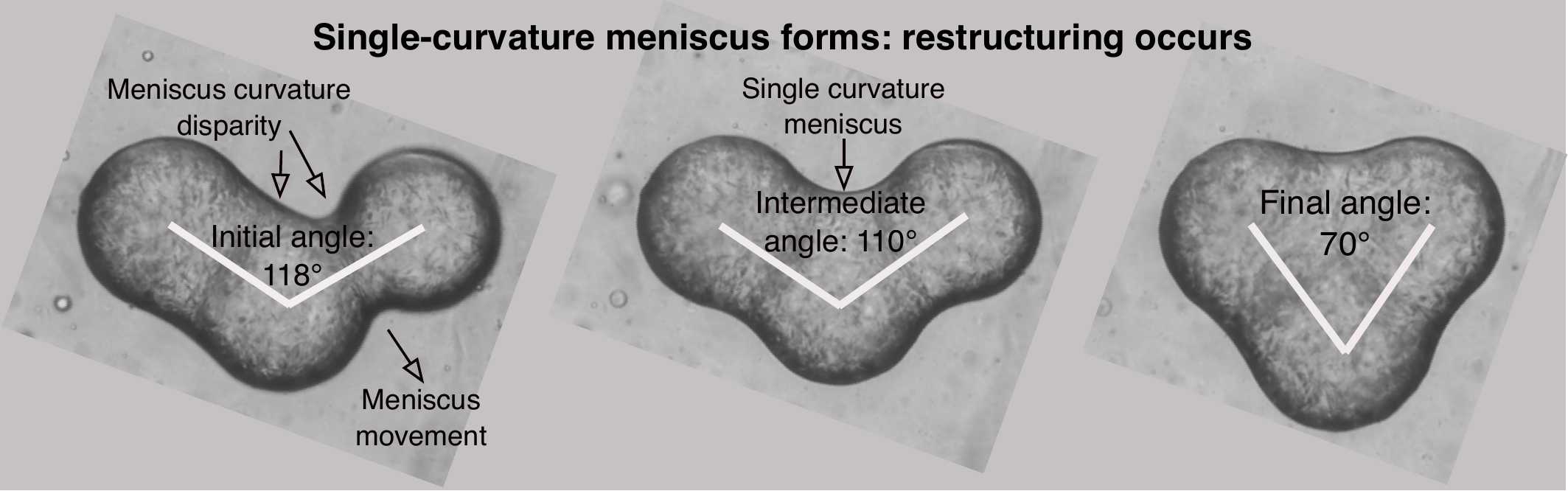}
  \caption{Close-up study of triplet formation in Figure \ref{fig:25restr}, showing neck 
width growth and the merging of the two separate meniscus regions, 
resulting in a significant amount of restructuring..}
  \label{fig:restrschem}
\end{figure*}

 \begin{figure*}
  \centering
  \includegraphics[scale=0.9]{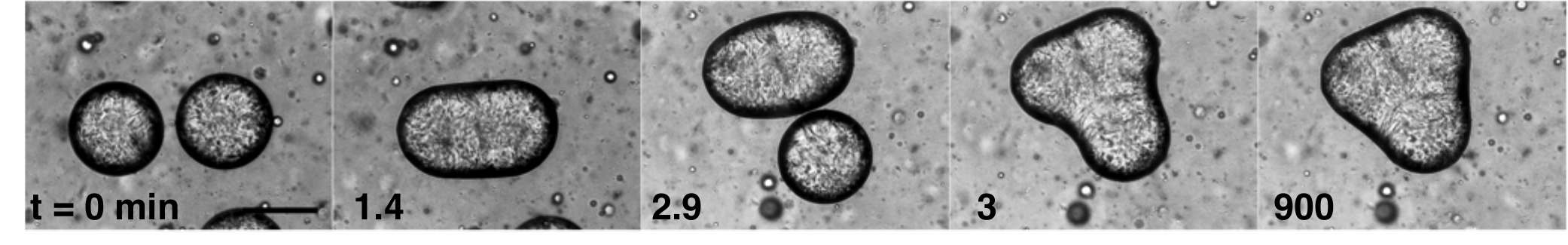}
  \caption{Restructuring of droplets (20\% solids level), showing more extreme changes in area and compactness of the triplet shape.
  Time in minutes, scale bar $200 \; \mu m$.}
  \label{fig:20restr}
\end{figure*}

\section{Experimental}

\subsection {Materials and methods}
Oil-in-water (O/W) emulsions are prepared by mixing 5 mL volumes of oil and aqueous phase. 
The dispersed oil phase is a mixture of hexadecane, 99\%, Sigma Aldrich, and petrolatum, 
Unilever, in a ratio yielding the desired solids concentration. The aqueous continuous phase 
is a neutralized 0.3 wt\% Carbopol in a 10 mM sodium dodecyl sulfate, SDS, 99\%, Fluka, 
surfactant solution. The emulsion is prepared by mixing dispersed and continuous phase, then 
heating to $75^\circ C$ and gently shaking for 10 seconds. A non-uniform emulsion size distribution is formed, 
which remains dispersed and suspended due to the yield stress of the continuous phase. 
Initially, the emulsion droplets are a homogeneous liquid phase, but as the temperature 
reaches the wax melting point of $\sim 40-60^\circ C$, elongated, flat, prismatic 
crystals form an elastic network inside the droplets, enabling arrest. Emulsions with 
different solid content, 15-45\%, are prepared to explore the behavior of various partially crystalline droplets.

\subsection {Microscopy}
The doublets and triplets are formed and studied by micromanipulation 
experiments.\cite{Pawar:2011tq,Pawar:2012wj,Dahiya:2016hz,thiel2016coalescence} 
A pre-pulled microcapillary is used to hold a droplet at its tip and then bring it 
into contact with another droplet to start the coalescence process. Borosilicate 
glass capillaries with1 mm OD and 0.5 mm ID, Sutter Instruments, are pulled 
with a Sutter Instruments Model P-97 Micropipette Puller to form microcapillaries 
with tips smaller than the droplets under study. The capillary is attached to a small 
water reservoir by Tygon tubing. Adjusting the height of the water reservoir 
varies the hydrostatic pressure, enabling the capillary to grab the droplet and  
manipulate it within the emulsion using a 3-axis coarse manipulator, Narishige 
International USA, mounted on a Motic AE31 inverted microscope.

An emulsion sample is placed on a glass slide and the tip of the capillary is aligned 
to the droplet under study while viewing through the microscope. Negative hydrostatic 
pressure is applied using the water reservoir to grip the droplet and then the manipulator 
used to move the drop across the continuous phase toward a second, and then third 
droplet to initiate coalescence, observe arrest, and document any restructuring that occurs. 
Because some structural relaxation occurs slowly, the shapes formed are observed 
continuously for 15 minutes and then compared for up to 15 hours after formation.

All results are for droplets that have been brought together in the level focal
plane of the microscope so no out of plane movements can bias the results. As the starting
doublets are rotationally symmetric about their long axes, our results should apply
to all starting orientations of the third droplet.

\section{Results and discussion}
Figure \ref{fig:multiangles} shows an example of a multidroplet complex 
emulsion microstructure exhibiting varying degrees 
of arrested coalescence and droplet arrangements.  Such structures are common in dairy food products 
where fat blends possess both solid and liquid components, providing the elasticity and interfacial forces
needed for arrest to occur.\cite{Pawar:2012wj}  Lines drawn over the image
in Figure \ref{fig:multiangles} indicate the orientation between droplet centers of mass
and map the wide variation in packing efficiency possible in an emulsion where arrest
is significant.  Some droplets form triangular packings with $\sim 60^\circ$ angles between them, 
a strong state based on hexagonal close packing, one of the 
densest possible arrangements of spheres. Others have much larger angles between them, indicating
that arrest stabilized the droplet arrangement before it could reach close packing.  
In an effort to understand the formation of similar multi-droplet structures, 
here we expand on previous studies of two-droplet structures
and investigate the effects of orientation on the final arrested state and packing efficiency of viscoelastic droplets.
 The goal is to better predict how structures like those in Figure  \ref{fig:multiangles} form and how
 they can be controlled and predicted.
 
\subsection{Elasticity-dominated structures}
In Figure \ref{fig:40restr}, droplets containing 40\% solids are brought into contact via 
micromanipulation, initiating coalescence and demonstrating the arrest of coalescence 
before it can complete formation of a larger spherical droplet.  The droplets possess a 
significant elasticity, with a $G' \sim\; 3\; kPa$, so a low arrest strain is expected\cite{Pawar:2012wj} 
and found to be less than $\epsilon \sim 1\%$ in  Figure \ref{fig:40restr}. The neck 
connecting the two droplets in the third frame of Figure \ref{fig:40restr} is about $50 \; \mu m$, 
or about 20\% of a single drop diameter, across because the high elasticity resists 
compression and exudes only a small amount of free oil to form the neck. This is sufficient, 
however, to join the droplets and arrest the doublet in a long-term stable nonspherical shape.\cite{Pawar:2012wj} 

Adding a third droplet requires formation of a second neck, necessitating further 
compression of the initial doublet to free up the fluid volume needed. We see this 
occurs in frames 4 and 5 of  Figure \ref{fig:40restr}, with the strain increasing slightly to 1.2\% 
and the neck width more than doubling to $117 \; \mu m$. Clearly the droplets 
are sufficiently adaptable to adjust and accommodate the addition of a new droplet 
as 60\% of its mass is free liquid that can flow out of the porous microstructure during 
compression, like a saturated sponge. A fascinating aspect of this result is the ability 
of the drop aggregate to adjust to numerous shape configurations depending on the 
environmental conditions and interactions.  An example of this was seen previously 
when the strain of a doublet oscillated between two deformation states because of changes in the 
local interfacial tension,\cite{Dahiya:2016hz} demonstrating significant degrees of 
reversibility and responsiveness.\cite{Caggioni:2014kw,Caggioni:2015kh} 
 
A key feature of the formation of the stable triplet in Figure \ref{fig:40restr} is the dynamic
behavior of the liquid neck bridging the droplets. It is instructive to examine in closer detail
what sorts of changes occur to understand the driving force behind structure formation. 
 Figure \ref{fig:norestrschem} shows a close-up view of frames 4 and 5 of Figure \ref{fig:40restr},
 allowing observation of the liquid neck meniscus and its shapes at different stages of the process.
 Although the free liquid permeates the entire triplet, some of its greatest mobility occurs at the 
 neck between droplets. We see in Figure \ref{fig:norestrschem} the meniscus transitions
 from a very small radius and a very highly curved state to a larger neck with a smaller curvature as the third
 droplet equilibrates into its final position. The two neck menisci remain distinct, with opposite curvature sign
 from the middle droplet surface, whereas full coalescence would likely have merged the two menisci
 and bridged the middle droplet. As a result, here the third droplet does not reorient because of the 
 meniscus expansion, its final angle through the center droplet is quite close to when coalescence began.
 If we now adjust the mechanical properties of the droplets, by decreasing solids level, we expect similar changes 
to those seen in Figure \ref{fig:40restr} but with increasing dominance of the interfacial driving force.

\subsection{Interface-dominated structures}
 For the case of droplets containing 30\% solids, we see in Figure \ref{fig:30restr} similar behavior to 
 Figure \ref{fig:40restr}, but with higher strains and neck widths. This is consistent 
 with a lower solids level and elasticity.\cite{Pawar:2012wj}
 Also noticeable is the merging of the two neck menisci on the inside edge of the triplet, 
 something that could not occur for the higher solids content case in Figure \ref{fig:40restr}. 
 Despite the formation of a single, lower curvature, meniscus from two higher curvature 
 menisci on one side of the triplet, the overall triplet structure changes only a small amount 
 due to restructuring. As in Figure \ref{fig:40restr}, the strain of each connected droplet pair does increase with time, 
 reflecting a similar need for additional compression to accommodate the creation of a second neck.
 Clearly a reduction in droplet elasticity lowers the resistance to interfacial pressures, enabling
 greater degrees of compression of the droplet microstructure and the beginning of restructuring behavior
 that alters the shape of the final triplet from its beginning orientation. Since such behavior has the potential
 to significantly vary emulsion structure and rheology, we further investigate the restructuring by
 altering the balance between the two driving forces more in favor of interfacial dominance: by examining
  even lower elasticity systems.

Figure \ref{fig:25restr} shows the coalescence of droplets containing 25\% solids, 
and they exhibit a similar initial arrest behavior to the higher solids cases in 
Figures  \ref{fig:40restr} and  \ref{fig:30restr}. Here the lower solids level and 
elasticity enables an even larger strain, $\epsilon \sim 10\%$, and thus a greater 
degree of meniscus expansion than the previous two concentrations. In Figure \ref{fig:25restr} 
the meniscus between the first and second droplet merges with the one 
between the second and third droplet between 1.3 minutes and 1.5 minutes, and the 
combined meniscus then continues its outward expansion. 
Between 1.5 and 2.5 minutes the fluid of the meniscus contains almost no solids, 
emphasizing the viscoelastic nature of these structures. After 2.5 minutes, 
Figure \ref{fig:25restr} shows a second way that addition of a third droplet dynamically 
modifies the droplet aggregate state: the third droplet is steadily pulled into a more 
compact packing through $\sim 50^\circ$ of motion. Examining the restructuring transition in 
more detail in the close-up study in Figure \ref{fig:restrschem}, we observe the importance 
of meniscus dynamics during triplet formation. Compression of the internal droplet solid 
network frees additional fluid to expand the menisci until they merge into one, single-curvature,
interface. The third droplet then moves around the assembly until it reaches a lower angle
with a much lower surface area and energy, while all meniscus curvatures have been reduced
to a more uniform value throughout the triplet. 

Such a change is significant: the initial state of the droplet connection would have, if retained as for the triplets in 
Figures \ref{fig:40restr} and \ref{fig:30restr}, created a more open, high fractal dimension, 
structure in a larger emulsion network. The meniscus-driven restructuring is a mechanism 
unique to this sort of system, balancing interfacial driving forces with elastic response and 
resulting in a strong means of actuating shape change. The movement represents a distinct 
pathway to structure formation in emulsion-based products. Typically emulsion networks are 
assumed to form via fractal aggregation models, depending on the significance of transport 
rates and droplet interactions. However, a more compact structure is shown here to be possible 
as a result of restructuring  of droplet aggregates, depending 
on the balance of elastic and interfacial forces. Such changes are expected to produce structures 
with varying degrees of compactness greater than expected for fractal systems.
More compact droplet arrangements are expected to impact emulsion mechanical properties, 
given the strength imparted by constructing a network from compact triangular building blocks.\cite{2012Hsiao}  
The observation that the restructuring occurs at an intermediate solids 
level indicates a need to understand the balance between the interfacial and elastic forces in these systems, 
so we continue to reduce the solids level in order to explore the system extremes.

Creating a droplet with 20\% solids continues the progression observed above of increasing dominance of 
interfacial forces over elastic resistance to deformation. The first two frames of Figure \ref{fig:20restr} 
show the construction of a doublet with a significantly higher strain than seen in 
Figures \ref{fig:40restr}-\ref{fig:25restr}. The lower elasticity of the droplets 
allows them to compress so far that the neck width is the same as the individual droplet diameter. When a
third droplet is added to produce a triplet we observe restructuring similar to Figure \ref{fig:25restr}, 
but at a more rapid pace than seen for 25\% solids in Figure \ref{fig:25restr}. Here the movement of 
the third droplet resembles more of a flow than the rigid body rotation seen in Figure \ref{fig:25restr}, 
as the lower elasticity of the droplet prevents it from remaining in its initial orientation and shape. 
The large neck and flow-like behavior of the restructuring suggests the droplet locally yields under the interfacial pressures
while the higher curvature ends of the droplet indicate that regions of elastic structure remain.

Comparing the final state of the triplets formed in Figures \ref{fig:40restr}-\ref{fig:20restr} 
demonstrates the increasing dominance of the liquid meniscus, transmitting interfacial pressure, in determining the 
final shape of the droplet aggregate. Increasing solids level increases elasticity and better 
preserves the individual spherical identity of the droplets.  All experiments here have examined 
the behavior of triplets formed in a single level plane of the surrounding fluid, allowing us to avoid any out-of-plane 
movement and track restructuring as it occurs for each droplet. Even with this simplified approach, 
the complexity of a triplet means there are a number of possible initial conditions and we would like to 
understand the variability of the restructuring phenomena observed for low-to-intermediate solids concentrations. A critical variable is 
expected to be the angle of approach by the third droplet, defined here as the angle the third drop's center of mass 
makes with the long axis of the doublet, as in Figures \ref{fig:norestrschem} and \ref{fig:restrschem}. Varying this angle provides 
a way to study the relevant processes and forces involved in restructuring: movement of a droplet initially attached at a high
angle of approach to a more compact packing arrangement.

  \begin{figure*}
  \centering
  \includegraphics[scale=0.5]{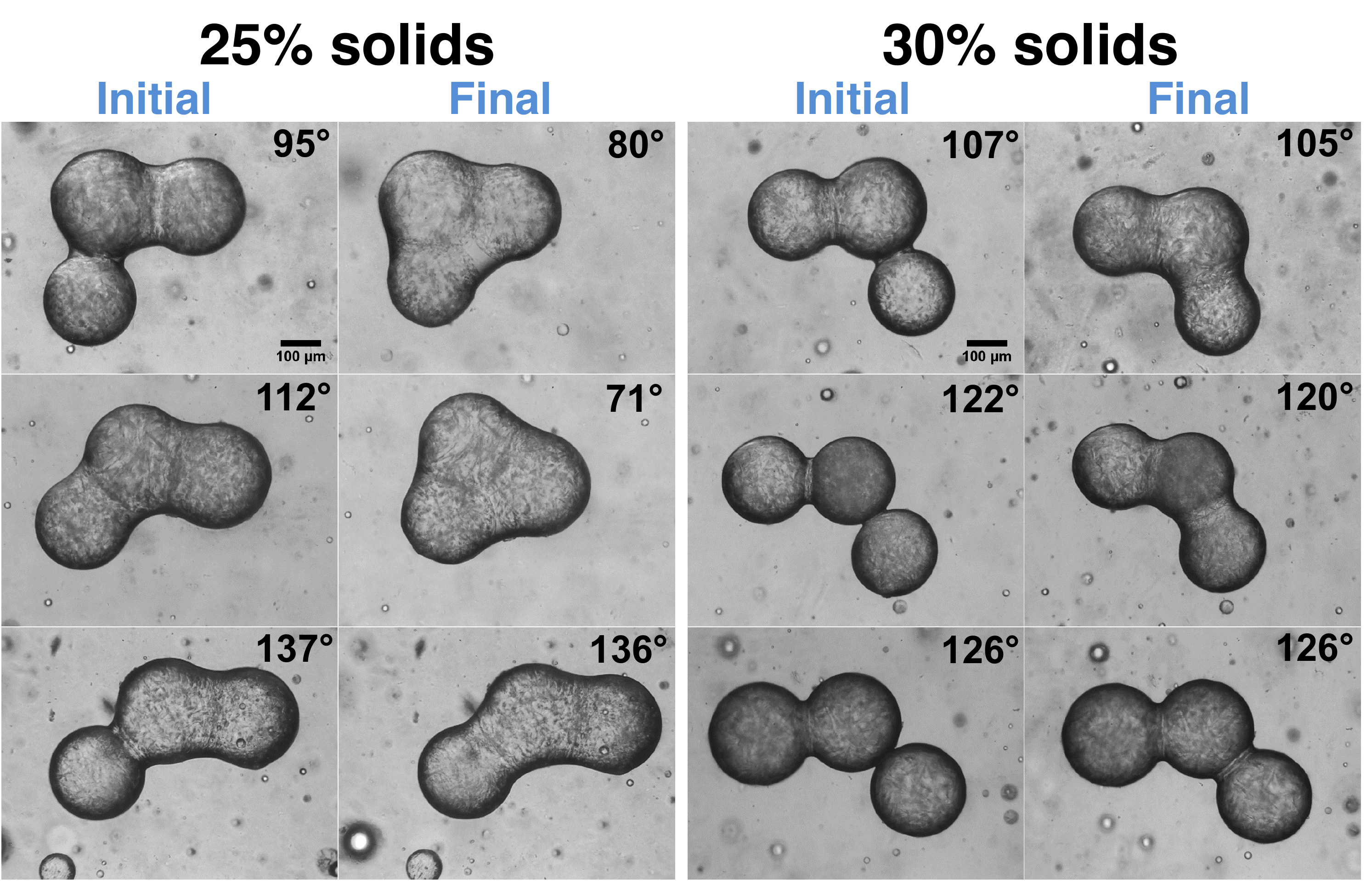}
  \caption{Images showing the effect of different approach angles on the restructuring of
   droplets containing 25\% and 30\% solid concentrations.}
  \label{fig:pics2530}
\end{figure*}

\subsection {Approach angle effects}
Observations of droplet coalescence during triplet formation in Figures \ref{fig:40restr}-\ref{fig:20restr} 
indicate the dynamic character of the triplet as a result of movement directed by the meniscus between drops. We hypothesize 
that the approach angle determines the degree to which the meniscus can alter the 
 droplet from its initial position via restructuring. Experiments varying the approach angle for all four solids concentrations 
were carried out in order to determine the effect of the initial droplet orientation on its final state. 
Examples of several different initial angles are shown
for 25\% and 30\% solids concentrations in Figure \ref{fig:pics2530}, along with the final angle attained after sufficient 
time allowed for restructuring.   For the 25\% solids case 
 in Figure \ref{fig:pics2530}, the approach angles vary between $95-137^\circ$ and only the highest 
 approach angle avoids restructuring. For the case of the lower two angles, the third droplets are seen 
 to be repositioned in the final state at an angle $25-52^\circ$ lower, respectively, than their initial state. 
 Between an approach angle of $112^\circ$ and $137^\circ$  there appears to be a threshold approach angle
  where the interfacial force exerted by the meniscus is unable to initiate, or significantly alter, the third droplet's
 orientation and the initial orientation remains stable. Interestingly, the two neck menisci overlap for $137^\circ$
 but the droplet does not restructure.  Clearly some form of resistance exists and could be related to the
 increased compression droplets may experience at larger approach angles.  At higher angles
 the interfacial force is directed more through the center of the droplet and may create local connections, or
 bonds, as a result of compressive stresses versus cases when the meniscus force is directed off of the center axis.
 
  A slight increase in solids concentration to 30\% mostly stabilizes the droplets against any
 restructuring, even for a similar angular range where significant restructuring is observed for 25\% solids in
 Figure \ref{fig:pics2530}. If the fluid meniscus joining these droplets is responsible
 for transmitting the force that causes restructuring, the observed dependence on approach angle
 may occur because higher approach angles decrease the volume of meniscus in one region. 
 The effect of solids level  likely occurs because increased
 elasticity will decrease compressibility of the internal crystalline microstructure, making less free fluid available at
 higher solids levels and also reducing meniscus volume and ability to restructure.  
 Following this reasoning, we examine the effect of approach angle for a wide range of solids concentrations
 in order to more closely map the boundary between stability and restructuring behavior and to understand its origin.

 \begin{figure*}
  \centering
  \includegraphics[scale=0.5]{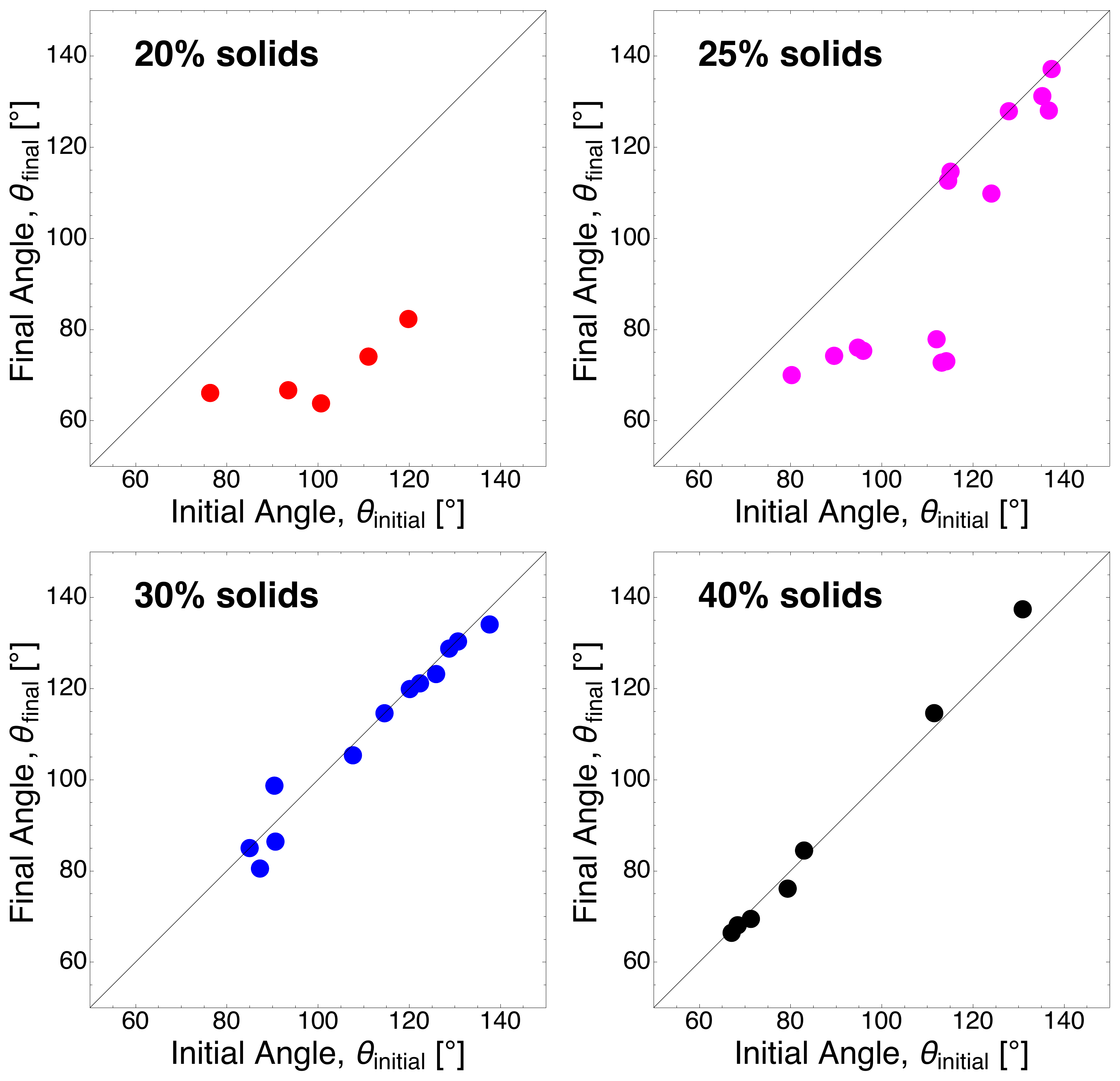}
  \caption{Results plotted for numerous triplet formation processes, mapping the initial 
  and final angle between the third droplet added and the initial doublet. The diagonal 
  line of equality indicates the cases when no restructuring occurs and deviation from 
  the line maps the boundaries of the specific cases of restructuring.}
  \label{fig:20253040angle}
\end{figure*}

Figure \ref{fig:20253040angle} shows plots of initial and final approach angle for four 
solids concentrations, with a solid line drawn for $\theta_{initial} = \theta_{final}$ to allow 
comparison of cases where restructuring occurs and the data deviate from the line.
At 20\% solids we see that all triplets exhibited restructuring. The only change is a slight increase in final angle
as approach angle is increased.  This may indicate an increased resistance to restructuring at higher 
initial angles with some of the droplet mass spreading out as it flows into its final position. Similarly, at
low initial angles the system converges to the state of most compact packing.
Increasing solids to 25\% we see a clear transitional behavior depending on approach angle.
Restructuring occurs for lower approach angles, up to a value $\sim 115^\circ$, 
above which the droplets abruptly no longer change. We also see the final angle of the 
restructured systems, at solids levels of 20\% and 25\%, 
tends to level off near a value of $70^\circ$. The consistent value reflects the 
approach to a true spherical close-packed value of $60^\circ$, differing only because the 
droplets' internal elasticity enables varying deformation, and some drag may reduce complete rearrangement.
For solids levels of 30\% and 40\% no restructuring is observed for any angle of approach and 
points in Figure \ref{fig:20253040angle} all lie on the equality line. There are some outliers
for 30\% and 40\% solids, where the final droplet angle is actually a bit higher than its starting value.
Presumably these rare effects occur because of some bias in the crystalline microstructure
that shifts the meniscus-driven movement of a third droplet away from the
typical direction of restructuring into a more compact shape.   
We have observed that an angle of approach exists for intermediate level solids concentrations
 that determines whether restructuring occurs. The value of this angle is $ \theta_c \sim 115^\circ$, from Figure  \ref{fig:20253040angle},
and may be a result of the relative location of the two menisci joining the outer two droplets 
to the center one in a triplet. 

\subsection{Triplet restructuring theory}
In order to better understand the transition angle at which triplets begin to restructure, as 
in Figures \ref{fig:25restr} and \ref{fig:restrschem}, we develop a simple theoretical model
of the phenomenon. Assuming that the primary driving force is interfacial
tension, the restructured triplet should be the lowest energy state
given its relatively small surface area. However, we assume that the
triplet is capable of restructuring to achieve this state only if
the two oil menisci, which bridge from the outer droplets to the center
drop, overlap. We assume that the oil bridge between droplets will
take the shape of a catenoid, which is known to be a minimal surface
between two loops that share an axis. Thus, a system of two droplets
connected by an oil bridge can be represented in two dimensions by
two circles connected by a catenoidal curve. The outer edges of two
droplets whose centers lie on the $x$ axis are,
\begin{eqnarray}
r_{0}^{2} & = & x^{2}+y_{d,0}^{2}\\
r_{1}^{2} & = & \left(x-\left(1-4\epsilon_{1}\right)r_{1}\right)^{2}+y_{d,1}^{2}
\end{eqnarray}
where $r_{i}$ is the radius of droplet $i$, and $\epsilon_{i}$ is
the strain between droplet $i$ and droplet $0$. The meniscus is written,
\begin{equation}
y_{m}=a\cosh\left(\frac{x-b}{a}\right)\,\,\,\,\,\,x_{0}\le x\le x_{i}
\label{eq:catenoid}
\end{equation}
where $a$, $b$, $x_{0}$, and $x_{i}$ are determined by assuming
that the meniscus and each droplet connect continuously and smoothly.

 \begin{figure*}
  \centering
  \includegraphics[scale=0.65]{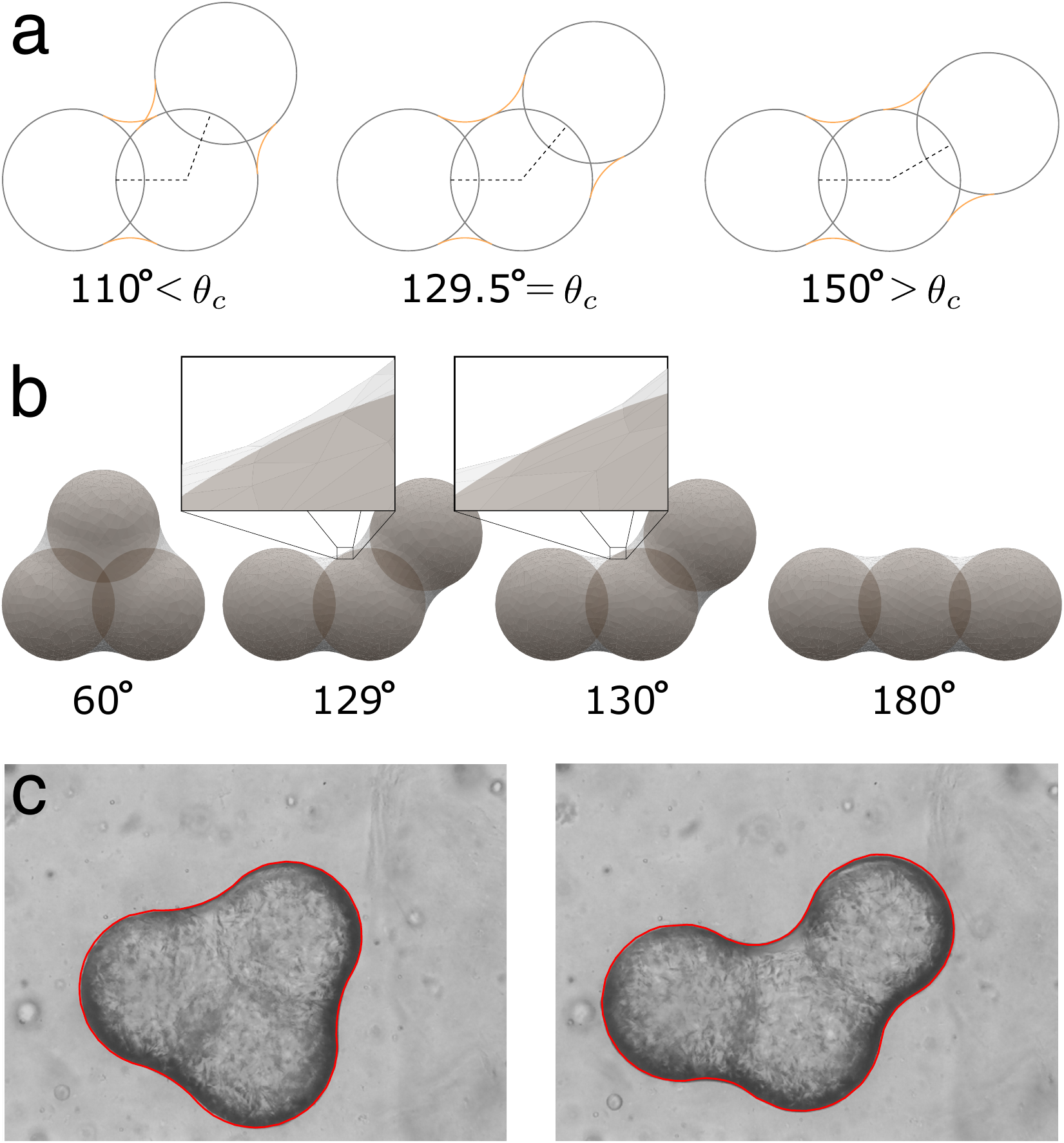}
  \caption{Panel a) shows diagrams of our analytical model for angles less than, equal to, and greater than
  the critical angle $\theta_c$ given equally sized droplets with a strain of 10\%.
  In the diagrams, a catenoidal meniscus (orange) forms between two each abutting pair of droplets (gray).
  The angle formed by the centers of two droplets and the point at which the meniscus meets the droplet exterior is $\theta_c/2$. 
  Panel c) shows a series of simulation visualizations at $\theta = 60, 129, 130$, 
  and $180 ^\circ $ for the series $\epsilon = 10\%$. The zoomed in frames demonstrate the 
  transition from overlapping to non-overlapping menisci that occurs at $\theta_c$ 
  between $129^\circ$ and $130^\circ$.
  Panel c) overlays the results of simulations (red outline) with experimental images for 25\%-solid triplets with 
  angles of $75^\circ$ (left) and  $115^\circ$ (right). The radii $r_i$ and stains $\epsilon_i$ 
  were extracted from the experimental images and used as inputs to the simulations.}
  \label{fig:theory1}
\end{figure*} 

This configuration appears in Figure \ref{fig:theory1}a for $r_{1}=r_{2}$ and $\epsilon_{1}=10\%$.
For each pair of abutting droplets, the points $x_{0}$ and $x_{i}$ determine the size of the meniscus between the droplets.
If the angle formed by the centers of mass of a triplet is less than some critical angle $\theta_c$,
the two menisci of the triplet will overlap and restructuring should occur. 

This analytical method provides an estimate of the transition angle
for restructuring for any given droplet sizes and strain amounts.
However, unlike the experimental system, this simple model does not
conserve the volume of oil available to form the meniscus. In order
to model a volume-constrained system with interfacial tension, we also perform simulations
using the \emph{Surface Evolver}.\cite{brakke1992surface}

With \textit{Surface Evolver}, we find a minimal surface subject to
the level set constraints,
\begin{eqnarray}
r_{0}^{2} & \ge & x^{2}+y^{2}+z^{2}\\
r_{1}^{2} & \ge & \left(x-R_{_{1,-}}\right)^{2}+y^{2}+z^{2}\\
r_{2}^{2} & \ge & \left(x+R_{_{2,+}}\cos\theta\right)^{2}+y^{2}+\left(z-R_{_{2,+}}\sin\theta\right)^{2},
\end{eqnarray}
and the volume constraint,
\begin{equation}
V=\frac{4}{3}\pi(r_{0}^{3}+r_{1}^{3}+r_{2}^{3}).
\end{equation}
Here $R_{_{i,\pm}}=\left(1-4\epsilon_{i}\right)r_{i}\pm r_{0}$ is the
distance from the origin to the center of sphere $i$, and $\theta$
is the angle formed by the centers of the spheres comprising the triplet.
Assuming interfacial tension is the dominant force that determines triplet
shape, the minimal surface given by \textit{Surface Evolver} should
be the shape of the oil in and around the droplets.

Panels b and c of Figure \ref{fig:theory1} visualize our volume-conserved simulations. 
In Figure \ref{fig:theory1}b, we highlight the point of 
disconnection of the menisci for simulations with $\epsilon=10\%$. 
Figure \ref{fig:theory1}c indicates that our simple interfacial tension-dominated 
model does very well to predict the shapes of the droplet triplets. 

\begin{figure*}
  \centering
  \includegraphics[scale=0.65]{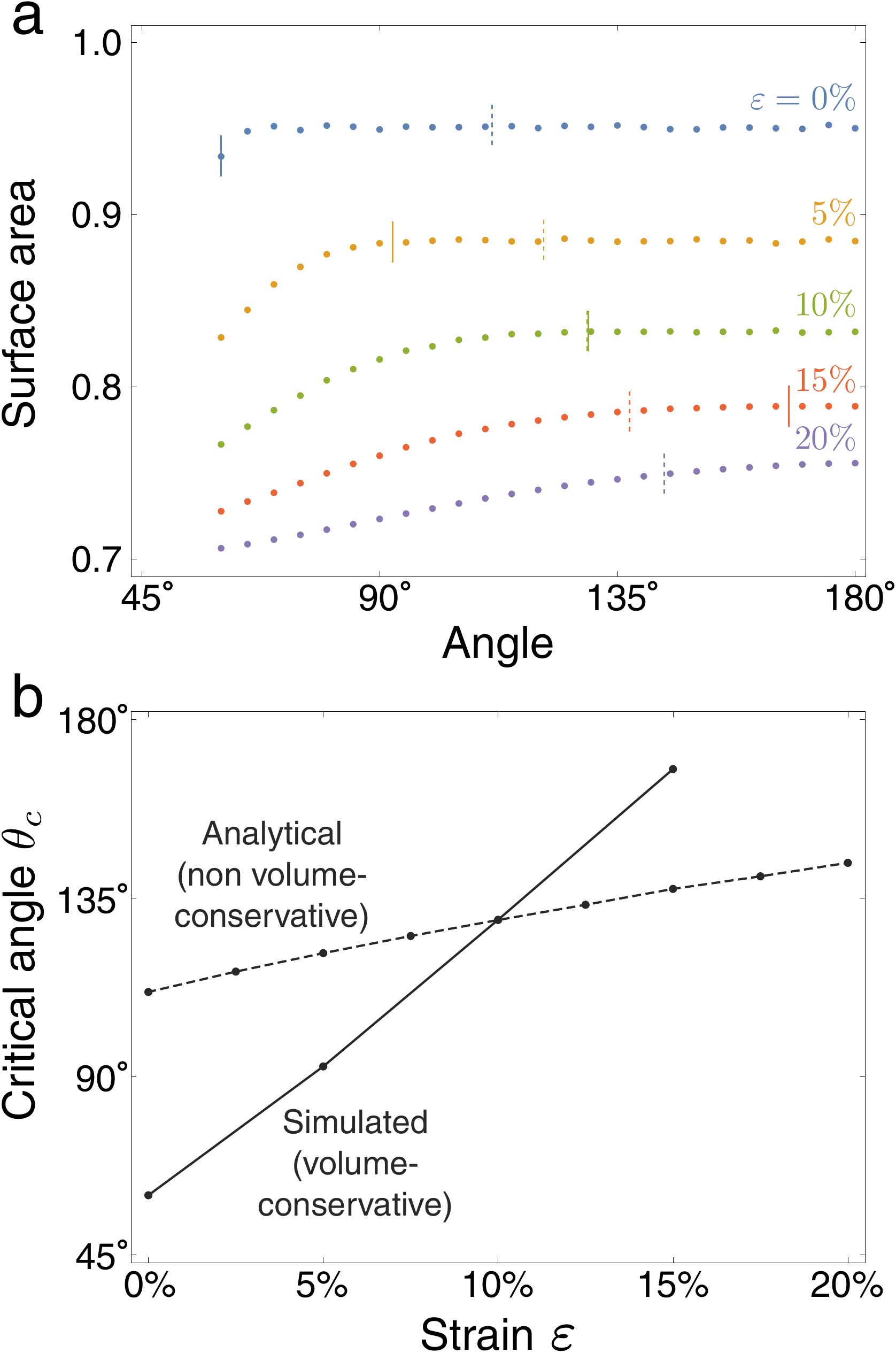}
  \caption{Panel a) shows surface area, normalized by the surface area of 
  three unconnected spheres, of several simulation runs for $r_0 = r_1 = r_2$. 
  The different series represent different values of strain $\epsilon$, 
  while the solid (dashed) vertical line in each series gives the critical transition angle based on simulation (analytical) results.
  Panel b) plots the critical angle $\theta_c$ as a function of strain based on analytical results (dashed line) which do not conserve droplet volume
  and simulation results (solid line) which do conserve droplet volume.
  No critical angle exists for simulated runs at $\epsilon=20\%$ because
  the volume available to form the menisci is large enough that the menisci are never disconnected.}
  \label{fig:theory2}
\end{figure*}

Figure \ref{fig:theory2}a shows the surface area of many simulated 
arrangements as a function of triplet angle and for several different strains. 
For each series, the minimum energy, which is proportional to surface area, is achieved at $60^\circ$, 
which corresponds to spherical close-packing. 
The energy of a triplet at a given angle then increases with 
increasing angle before leveling off to a constant interfacial tension. 
The leveling off occurs exactly at the critical angle $\theta_{c}$ that 
corresponds to the menisci becoming fully disconnected, 
as we theorized for the basis of our analytical model. In Figure \ref{fig:theory2}a 
the critical angles are noted by a vertical solid line on each data series, 
solid for the volume-conserving simulation and dashed for the analytical results.
Though second-order effects undoubtedly exist in this system, the simple 
approximation of an interfacial-tension-dominated set of overlapping spheres 
does well to model our experimental observations and confirm our physical intuition. 
The simulation is also consistent with the physical case, as lower solids 
levels have more overlap and greater amounts of free fluid available.

The effect of the volume constraint on the critical angle is 
evident from Figure \ref{fig:theory2}b. Simulations with little strain
have less oil available and therefore form narrow menisci that are
less likely to overlap. Thus, the simulated results show a critical
angle that is shifted below the angle predicted by the analytical
construction that does not conserve volume. 
This helps to explain why restructuring is less common for stiffer droplets.
Similarly, simulations with large overlap have more meniscus material
available and thus exhibit a shift of the critical angle towards larger
values than that predicted by non-conservative theory. For $\epsilon=10\%$,
the analytical model does coincidentally conserve volume, and no shift
in the critical angle is observed.

That we do observe a critical angle experimentally suggests that,
even though restructuring is always favorable given the global energy
landscape, the triplet is only capable of restructuring if the derivative
of interfacial tension with respect to angle overcomes some initial energy
cost \textemdash{} likely due to mechanical friction. This explains
the discrepancy between the critical angles observed experimentally
and found by simulation. Based on experimental evidence, droplets with 25\% solids
tend to have overlaps between $\epsilon=10\%$ and $\epsilon=15\%$, corresponding
to the third or fourth series from the top in Figure \ref{fig:theory2}a,
which suggests a critical angle of at least $129^\circ$ and not $115^\circ$
 as observed. Though the interfacial tension produces some torque
to restructure the triplet, the triplet only restructures if this
torque overcomes the mechanical friction opposing it.  This work has not explicitly examined the
effects of droplet size but the fact that the restructuring behavior seems
to depend on a balance between interfacial and frictional forces indicates
the phenomenon will be most important for droplets with sizes equal to, or smaller than, those
studied here.

\begin{figure*}
  \centering
  \includegraphics[scale=0.4]{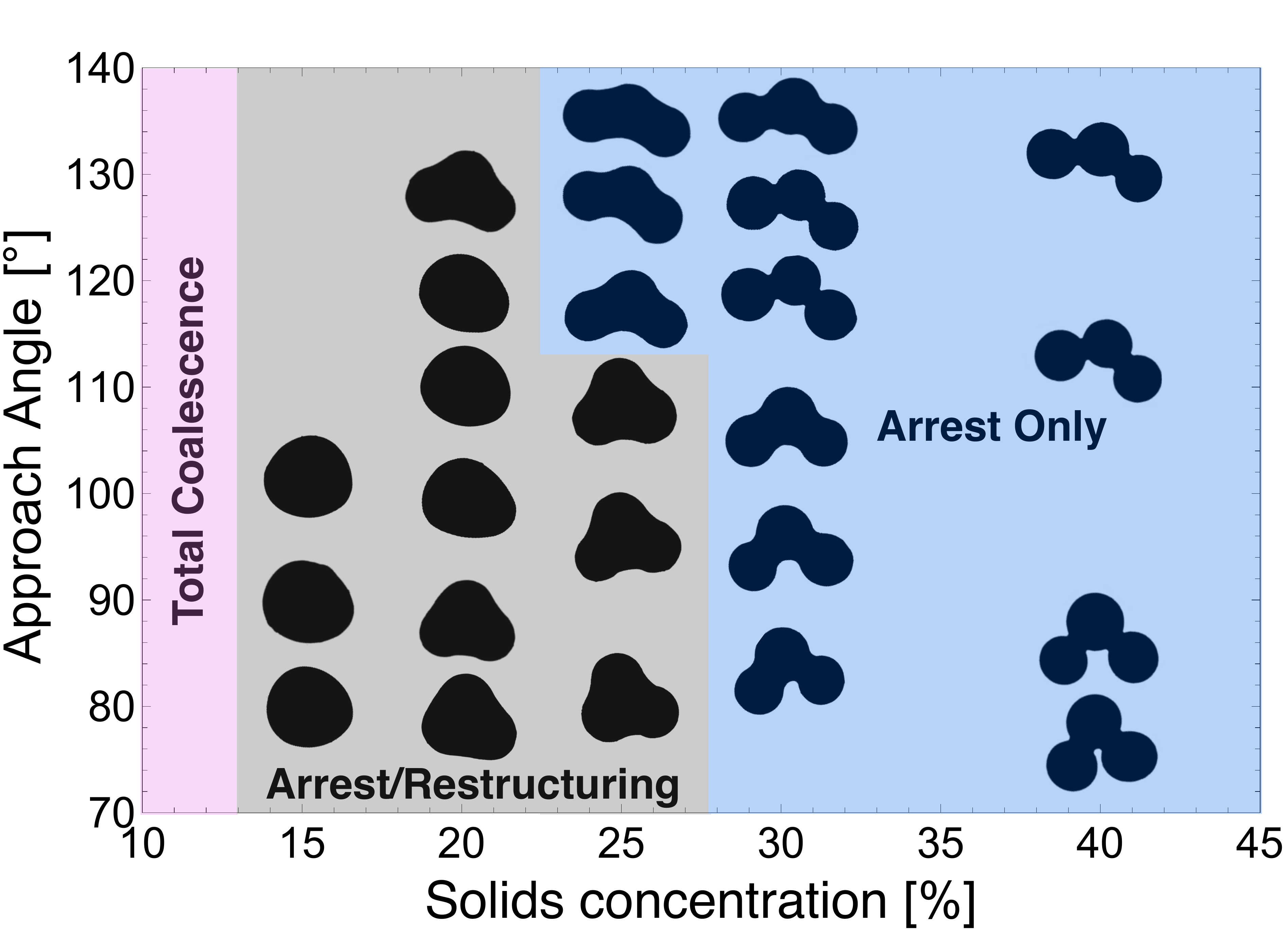}
  \caption{A design map of the possible shapes formed following the initiation of 
  coalescence in an aggregate of three viscoelastic droplets. The shapes are a result of the balance 
  between interfacial pressures and elastic deformation of droplet microstructure and 
  indicate where shape complexity, stability, and dynamic behavior is found.}
  \label{fig:shape}
\end{figure*}

Figure \ref{fig:shape} shows a map of the main modes of arrested coalescence behavior of the 
 triplets studied here. Consistent with earlier observations, we see that three regions 
are possible: total coalescence and loss of individual shape, 
arrested coalescence with subsequent restructuring, and
arrest without restructuring. Restructuring is possible where approach angles and elasticity are low, while sufficiently
high approach angle or elasticity can prohibit restructuring and arrest.  Some variability is noted in the observed trends,
such as when we see merging of menisci between droplets but no significant restructuring, Figure \ref{fig:30restr}.
Such variations indicate triplets can vary in their resistance to restructuring, likely due to small differences in crystal 
microstructure and mechanical properties.

Key example silhouettes of numerous final triplet shapes are also plotted in Figure \ref{fig:shape}
as a function of their solids concentration and initial angle of approach. 
The mapping of shapes reveals the regions of concentration and trajectory where droplet 
geometries are stable or forbidden based on the balance of interfacial and elastic energies.
One interesting effect is during approach angles of $100-125^\circ$ for 
20\% solids: the system passes through a local maximum in surface-to-volume
ratio with more compact shapes produced. Initially surprising, we assume this results from a 
change in the displacement possible for the third droplet: the longer the distance it is pulled 
the more it is fluidized and spread/compacted during interface equilibration, producing a more compact shape.
At high or low approach angles the droplet travels almost no distance from its original contact point
and less deformation occurs.

\section{Conclusions}
Structural dynamics in aggregates of three viscoelastic droplets have been studied during 
arrested coalescence in order to understand the changes in microstructure 
wrought by interfacial and elastic driving forces in an emulsion. The 
experimental data depict the change in deformation of droplet doublets when 
a third droplet is added to the system. A central player is the meniscus of fluid 
between droplets as it connects the droplets and exerts the shaping and 
equilibrating force of the interfacial tension on the aggregate. The solids 
content of the droplet exerts an elastic opposition to the interfacial pressure 
and the ultimate balance between forces determines the final state of the 
droplet. There are two key effects observed for triplets that are unique from 
the doublet case and can be explained with simple physical arguments. 
The first is an increased width of the fluid neck bridging droplets that is a direct 
result of the redirection of liquid to create a second neck and connect the third droplet.  
A second behavior has an even more profound effect on the droplet network structure and 
packing: the relocation of the third droplet to pull it into a more close-packed state and 
restructure the triplet from its initial position. None of the resulting structures are fully coalesced 
and their arrested state is a key factor in their formation, but this work shows that there 
are additional relevant dynamics in this process than have been previously considered 
in the formation of three-dimensional emulsion structures. These observations are 
important in real emulsions as the flow trajectory and droplet addition history become 
particularly important to the final emulsion structure. Further research is needed to see 
the effect of restructuring in a system with larger, more realistic, numbers of droplets in 
bulk emulsions.  These results add to our physical model of arrested coalescence of 
droplets, moving us closer to the simulation of the three-dimensional networks formed 
during mixing and preparation of foods, pharmaceuticals, cosmetics, and coatings with 
targeted rheological properties. The restructuring mechanism also indicates a method to 
program varying shape formation and shape change into colloidal systems from generic 
starting conditions, adding to the symmetries possible in assemblies constructed from 
fully solid particles by a liquid interface.\cite{Manoharan:2003vj,Meng:2010gsa,Kraft:2009vf, Arkus:2009dc} 
The restructuring studied here is a complex mix of deformation and movement of 
connected droplets so, although we know interfacial tension is the driving force for the 
process, additional measurements of the mechanism and magnitude of 
resistance to restructuring are still needed. 

\vspace{5pt}

\noindent{\bf Acknowledgements:}  PD and PTS gratefully acknowledge financial support from the Procter \& Gamble 
Company, the 2014 UNSW Major Research Equipment Infrastructure Initiative, 
and ARC Discovery Grant DP150100865. TJA and ADB were funded by a Tufts 
International Research Grant to perform part of the work at UNSW. TJA is supported 
by a Cottrell Award from the Research Corporation for Science Advancement.

\bibliographystyle{apsrev4-1}
\bibliography{triplets}

\end{document}